# Effect of Tensor Interaction in the Dirac-Attractive Radial Problem under Pseudospin Symmetry limit

M. Hamzavi[1*], M. Eshghi[2], S. M. Ikhdair[3]

[1]*Department of Basic Sciences, Shahrood Branch, Islamic Azad University, Shahrood, Iran*

[2]*Department of Physics, Central Tehran Branch, Islamic Azad University Tehran, Iran*

[3]*Physics Department, Near East University, 922022 Nicosia, North Cyprus, Mersin 10, Turkey*

*Corresponding author: Tel.:+982733395270, fax: +982733395270*

**Email:** majid.hamzavi@gmail.com

## Abstract

We approximately investigated pseudospin symmetric solutions of the Dirac equation for attractive radial potential including a Coulomb-like tensor interaction under pseudospin symmetry limit for any spin-orbit quantum number $\kappa$. By using the parametric generalization of the Nikiforov-Uvarov method, the energy eigenvalues equation and the corresponding wave functions have been obtained in closed forms. Some numerical results are also given. We presented tensor interaction removes degeneracy between two states in pseudospin doublets.

## 1. Introduction

It is well known that the exact energy eigenvalues of the bound state play an important role in quantum mechanics. In particular, the Dirac equation, which describes the motion of a spin-1/2 particle, has been used in solving many problems of nuclear and high-energy physics. Recently, there has been an increased in searching for analytic solution of the Dirac equation [1-7].

Within the framework of Dirac equation, pseudospin symmetry used to feature deformed nuclei, superdeformation, to establish an effective shell-model [8-10] and spin symmetry is relevant for mesons [11]. The spin symmetry appears when the magnitude of the scalar and vector potentials are nearly equal, i.e., $V_v(r) \cong V_s(r)$, in the nuclei (*i.e.*, when the difference potential $\Delta(r) = V_v(r) - V_s(r) = C_s = const.$). However, the pseudo-spin symmetry occurs when $V_v(r) \cong -V_s(r)$ (*i.e.*, when the sum potential $\Sigma(r) = V_v(r) + V_s(r) = C_{ps} = const.$) [12]. The bound states of nucleons seem to be sensitive to some mixtures of these potentials. The $\Delta(r) = 0$ and $\Sigma(r) = 0$ correspond to *SU*(2) symmetries of the Dirac Hamiltonian [13].

Concepts of spin and pseudospin symmetries and a tensor potential have been found interesting applications in the field of nuclear physics [14-16]. On the other hand, tensor potentials were introduced into the Dirac equation with the substitution $\vec{p} \to \vec{p} - im\omega\beta \cdot \hat{r} U(r)$ [17, 18]. In this way, a new spin-orbit coupling term is added to the Dirac Hamiltonian. Recently, tensor couplings have been used widely in the studies of nuclear properties. In this regard, see [19-32].

The attractive radial potential, introduced by Williams and Poulios [33], is given by

$$V(r) = \frac{V_1 e^{-4ar} + V_2 e^{-2ar} + V_3}{(1 - e^{-2ar})^2} \tag{1}$$

where $V_1 = \alpha^2/4$, $V_2 = [(A-8)\alpha^2]/4$, $V_3 = [(4-A)\alpha^2]/4$, parameters $\alpha$ and $A$ are real and $\alpha > 1/2$, $4 < A < 8$. By using the concept of supersymmetry quantum mechanics, Zou et al. obtained exact energy equation of the Dirac equation for *s*-wave bound states of the attractive radial potential [34]. In figure 1, we plot this potential for various $\alpha$ and $A$. From figure 1, one can see that above potential has behavior like Coulomb attractive potential.

Tensor Coulomb-like potential [28] is

$$U(r) = -\frac{H}{r}, \qquad H = \frac{Z_a Z_b e^2}{4\pi\varepsilon_0}, \qquad r \geq R_c \tag{2}$$

where $R_c = 7.78\,fm$ is the Coulomb radius, $Z_a$ and $Z_b$ denote the charges of the projectile *a* and the target nuclei *b*, respectively.

Our aim in this paper is solve the Dirac equation for attractive radial potential including a Coulomb-like tensor coupling under the pseudospin symmetry limit. We obtain the energy eigenvalues equation and the corresponding spinor wave functions by using the parametric generalization of the Nikiforov-Uvarov (NU) method. We discuss the effect of tensor potential, too.

**2. Parametric Generalization of the NU method**

The NU method is used to solve second order differential equations with an appropriate coordinate transformation $s = s(r)$ [35]

$$\psi_n''(s) + \frac{\tilde{\tau}(s)}{\sigma(s)}\psi_n'(s) + \frac{\tilde{\sigma}(s)}{\sigma^2(s)}\psi_n(s) = 0, \tag{3}$$

where $\sigma(s)$ and $\tilde{\sigma}(s)$ are polynomials, at most of second degree, and $\tilde{\tau}(s)$ is a first-degree polynomial. To make the application of the NU method simpler and direct without need to check the validity of solution. We present a shortcut for the method. So, at first we write the general form of the Schrödinger-like equation (3) in a more general form applicable to any potential as follows [36]

$$\psi_n''(s) + \left(\frac{c_1 - c_2 s}{s(1-c_3 s)}\right)\psi_n'(s) + \left(\frac{-As^2 + Bs - C}{s^2(1-c_3 s)^2}\right)\psi_n(s) = 0, \tag{4}$$

satisfying the wave functions

$$\psi_n(s) = \phi(s) y_n(s). \tag{5}$$

Comparing (4) with its counterpart (3), we obtain the following identifications:

$$\tilde{\tau}(s) = c_1 - c_2 s, \quad \sigma(s) = s(1-c_3 s), \quad \tilde{\sigma}(s) = -As^2 + Bs - C. \tag{6}$$

Using the NU method, the eigenfunctions and eigenvalues of Eq. (4) are given by [36]

$$\rho(s) = s^{c_{10}-1}(1-c_3 s)^{\frac{c_{11}}{c_3}-c_{10}-1},$$

$$\phi(s) = s^{c_{12}}(1-c_3 s)^{-c_{12}-\frac{c_{13}}{c_3}},$$

$$y_n(s) = P_n^{(c_{10}-1, \frac{c_{11}}{c_3}-c_{10}-1)}(1-2c_3 s),$$

$$\psi(s) = s^{c_{12}}(1-c_3 s)^{-c_{12}-\frac{c_{13}}{c_3}} P_n^{(c_{10}-1, \frac{c_{11}}{c_3}-c_{10}-1)}(1-2c_3 s), \tag{7}$$

and

$$c_2 n - (2n+1)c_5 + (2n+1)(\sqrt{c_9} - c_3\sqrt{c_8}) + n(n-1)c_3 + c_7 + 2c_3 c_8 - 2\sqrt{c_8 c_9} = 0, \tag{8}$$

respectively, where the parametric constants are found as

$$c_4 = \frac{1}{2}(1-c_1), \qquad c_5 = \frac{1}{2}(c_2 - 2c_3),$$

$$c_6 = c_5^2 + A, \qquad c_7 = 2c_4 c_5 - B,$$

$$c_8 = c_4^2 + C, \qquad c_9 = c_3 c_7 + c_3^2 c_8 + c_6,$$

$$c_{10} = c_1 + 2c_4 - 2\sqrt{c_8}, \quad c_{11} = c_2 - 2c_5 + 2\left(\sqrt{c_9} - c_3\sqrt{c_8}\right),$$

$$c_{12} = c_4 - \sqrt{c_8}, \qquad c_{13} = c_5 - \left(\sqrt{c_9} - c_3\sqrt{c_8}\right). \tag{9}$$

In the rather more special case of $c_3 = 0$ [36],

$$\lim_{c_3 \to 0} P_n^{(c_{10}-1,\frac{c_{11}}{c_3}-c_{10}-1)}(1-2c_3 s\, 0) = L_n^{c_{10}-1}(c_{11} s) \tag{10a}$$

$$\lim_{c_3 \to 0}(1-c_3 s)^{-c_{12}-\frac{c_{13}}{c_3}} = e^{c_{13} s} \tag{10b}$$

and, from Eq. (5), we find for the wavefunction

$$\psi(s) = s^{c_{12}} e^{c_{13} s} L_n^{c_{10}-1}(c_{11} s) \tag{11}$$

## 3. Solution of the Dirac Equation including Tensor Coupling

The Dirac equation for fermionic massive spin-$1/2$ particles moving in an attractive scalar potential $S(r)$, a repulsive vector potential $V(r)$ and a tensor potential $U(r)$ is $[\hbar = c = 1]$

$$\left[\vec{\alpha}\cdot\vec{p} + \beta(M + S(r)) - i\beta\vec{\alpha}\cdot\hat{r} U(r)\right]\psi(\vec{r}) = \left[E - V(r)\right]\psi(\vec{r}), \tag{12}$$

where $E$ is the relativistic energy of the system, $\vec{p} = -i\vec{\nabla}$ is the three-dimensional momentum operator and $M$ is the mass of the fermionic particle. $\vec{\alpha}$ and $\beta$ are the $4 \times 4$ usual Dirac matrices [37, 38]. Following procedure Eq. (7) to Eq. (10) from Ref. [22], we have two coupled differential equations for lower and upper radial wave functions $G_{n\kappa}(r)$ and $F_{n\kappa}(r)$ as

$$\left(\frac{d}{dr} + \frac{\kappa}{r} - U(r)\right) F_{n\kappa}(r) = (M + E_{n\kappa} - \Delta(r)) G_{n\kappa}(r), \tag{13a}$$

$$\left(\frac{d}{dr} - \frac{\kappa}{r} + U(r)\right) G_{n\kappa}(r) = (M - E_{n\kappa} + \Sigma(r)) F_{n\kappa}(r), \tag{13b}$$

where

$$\Delta(r) = V(r) - S(r), \tag{14a}$$

$$\Sigma(r) = V(r) + S(r). \tag{14b}$$

Eliminating $F_{n\kappa}(r)$ and $G_{n\kappa}(r)$ from Eqs. (13), we obtain the following two Schrödinger-like differential equations for the upper and lower radial spinor components, respectively:

$$\left[ \frac{d^2}{dr^2} - \frac{\kappa(\kappa+1)}{r^2} + \frac{2\kappa}{r}U(r) - \frac{dU(r)}{dr} - U^2(r) + \frac{\frac{d\Delta(r)}{dr}}{M + E_{n\kappa} - \Delta(r)}\left(\frac{d}{dr} + \frac{\kappa}{r} - U(r)\right) \right] F_{n\kappa}(r)$$

$$= \left[ (M + E_{n\kappa} - \Delta(r))(M - E_{n\kappa} + \Sigma(r)) \right] F_{n\kappa}(r), \tag{15}$$

$$\left[ \frac{d^2}{dr^2} - \frac{\kappa(\kappa-1)}{r^2} + \frac{2\kappa}{r}U(r) + \frac{dU(r)}{dr} - U^2(r) + \frac{\frac{d\Sigma(r)}{dr}}{M - E_{n\kappa} + \Sigma(r)}\left(\frac{d}{dr} - \frac{\kappa}{r} + U(r)\right) \right] G_{n\kappa}(r)$$

$$= \left[ (M + E_{n\kappa} - \Delta(r))(M - E_{n\kappa} + \Sigma(r)) \right] G_{n\kappa}(r) \tag{16}$$

where $\kappa(\kappa-1) = \tilde{l}(\tilde{l}+1)$ and $\kappa(\kappa+1) = l(l+1)$. The quantum number $\kappa$ is related to the quantum numbers for spin symmetry $l$ and pseudospin symmetry $\tilde{l}$ as

$$\kappa = \begin{cases} -(l+1) = -(j+\frac{1}{2}) & (s_{1/2}, p_{3/2}, etc.) \quad j = l + \frac{1}{2}, \text{ aligned spin } (\kappa < 0) \\ +l = +(j+\frac{1}{2}) & (p_{1/2}, d_{3/2}, etc.) \quad j = l - \frac{1}{2}, \text{ unaligned spin } (\kappa > 0), \end{cases}$$

and the quasidegenerate doublet structure can be expressed in terms of a pseudospin angular momentum $\tilde{s} = 1/2$ and pseudo-orbital angular momentum $\tilde{l}$, which is defined as

$$\kappa = \begin{cases} -\tilde{l} = -(j+\frac{1}{2}) & (s_{1/2}, p_{3/2}, etc.) \quad j = \tilde{l} - \frac{1}{2}, \text{ aligned pseudospin } (\kappa < 0) \\ +(\tilde{l}+1) = +(j+\frac{1}{2}) & (d_{3/2}, f_{5/2}, etc.) \quad j = \tilde{l} + \frac{1}{2}, \text{ unaligned pseudospin } (\kappa > 0), \end{cases}$$

where $\kappa = \pm 1, \pm 2, \ldots$. For example, $(1s_{1/2}, 0d_{3/2})$ and $(1p_{3/2}, 0f_{5/2})$ can be considered as pseudospin doublets.

### 3.1. Pseudospin symmetry limit

In the pseudospin symmetry limit $\frac{d\Sigma(r)}{dr}=0$ or $\Sigma(r)=C_{ps}=$ constant [39, 40], then Eq. (16) with $\Delta(r)$ as attractive radial potential and a Coulomb-like potential tensor potential becomes

$$\left\{\frac{d^2}{dr^2}-\frac{\Lambda_\kappa(\Lambda_\kappa-1)}{r^2}-\tilde{\gamma}\left[\frac{V_1e^{-4\alpha r}+V_2e^{-2\alpha r}+V_3}{(1-e^{-2\alpha r})^2}\right]-\tilde{\beta}^2\right\}G_{n\kappa}(r)=0 \qquad (17)$$

where $\kappa=l$ and $\kappa=-l-1$ for $\kappa<0$ and $\kappa>0$, respectively, $\tilde{\gamma}=E_{n\kappa}-M-C_{ps}$, $\tilde{\beta}^2=(M+E_{n\kappa})(M-E_{n\kappa}+C_{ps})$ and $\Lambda_\kappa=\kappa+H$ is new spin-orbit quantum number. Above equation can not be solved analytically because of pseudocentrifugal term $\Lambda_\kappa(\Lambda_\kappa-1)/r^2$ term. Thus, we used the approximation scheme suggested by Greene and Aldrich [41] as

$$\frac{1}{r^2}\approx 4\alpha^2\left[C_0+\frac{e^{-2\alpha r}}{(1-e^{-2\alpha r})^2}\right], \qquad (18)$$

where the parameter $C_0=1/12$ is dimensionless constant [42]. Therefore, to see the accuracy of our approximation, we plotted the pseudocentrifugal term, $\frac{1}{r^2}$ and its approximation with parameter $\alpha=0.6$, in figure 2.

To obtain solution of Eq. (17), by using change of variables $s=e^{-2\alpha r}$, we rewrite it as follows

$$\left\{\frac{d^2}{ds^2}+\frac{1-s}{s(1-s)}\frac{d}{ds}-\frac{\Lambda_\kappa(\Lambda_\kappa-1)}{s^2}\left(C_0+\frac{s}{(1-s)^2}\right)\right.$$
$$\left.-\frac{\tilde{\gamma}}{4\alpha^2 s^2}\left(\frac{V_1 s^2+V_2 s+V_3}{(1-s)^2}\right)-\frac{\tilde{\beta}^2}{4\alpha^2 s^2}\right\}G_{n\kappa}(s)=0 \qquad (19)$$

Comparing Eq. (19) and Eq. (4), we can easily obtain the coefficients $c_i$ ($i=1,2,3$) and analytical expressions $A$, $B$ and $C$ as follows

$$c_1=1, \qquad A=\Lambda_\kappa(\Lambda_\kappa-1)C_0+\frac{\tilde{\gamma}V_1}{4\alpha^2}+\frac{\tilde{\beta}^2}{4\alpha^2}$$

$$c_2=1, \qquad B=\Lambda_\kappa(\Lambda_\kappa-1)(2C_0-1)-\frac{\tilde{\gamma}V_2}{4\alpha^2}+\frac{2\tilde{\beta}^2}{4\alpha^2}$$

$$c_3 = 1, \qquad C = \Lambda_\kappa(\Lambda_\kappa - 1)C_0 + \frac{\tilde{\gamma}V_3}{4\alpha^2} + \frac{\tilde{\beta}^2}{4\alpha^2} \qquad (20)$$

The specific values of coefficients $c_i$ ($i = 4, 5, ..., 13$) are found from Eq. (9) and displayed in table 1. By using Eq. (8), we can obtain the hole energy states as

$$\left(2n + 1 + \sqrt{(2\Lambda_\kappa - 1)^2 + \frac{\tilde{\gamma}}{\alpha^2}(V_1 + V_2 + V_3)} - \sqrt{4\Lambda_\kappa(\Lambda_\kappa - 1)C_0 + \frac{1}{\alpha^2}(\tilde{\gamma}V_3 + \tilde{\beta}^2)}\right)^2$$

$$= \frac{\tilde{\gamma}}{\alpha^2}(V_1 + V_2 + 2V_3) + \frac{\tilde{\beta}^2}{\alpha^2} + 4\Lambda_\kappa(\Lambda_\kappa - 1)C_0. \qquad (21)$$

Some numerical results are given in table 2. We use the parameters $M = 5.0\,fm^{-1}$, $C_{ps} = 0$, $H = 1$, $A = 5$ and $\alpha = 0.6$. For a given value of $n$ and $\kappa$, the above quartic equation, provides four distinct positive and negative real and complex energy spectra related with $E_{n\kappa}^+$ or $E_{n\kappa}^-$, respectively. Therefore, the procedures for calculating the four distinct energies; namely, $E_{n\kappa}^{(1)}$, $E_{n\kappa}^{(2)}$, $E_{n\kappa}^{(3)}$, $E_{n\kappa}^{(4)}$ are given, in details, in Appendix B of Ref. [43]. In tables 2, we consider the same set of spin doublets as: $(ns_{1/2}, (n-1)d_{3/2})$, $(np_{3/2}, (n-1)f_{5/2})$, $(nd_{5/2}, (n-1)g_{7/2})$, .... We see that the tensor interaction removes the degeneracy between two stats in spin doublets. When $H \neq 0$, the energy levels of the pseudospin aligned states and pseudospin unaligned states move in the opposite directions. For example: in pseudospin doublet $(1s_{1/2}, 0d_{3/2})$; when $H = 0$, $E_{1,-1} = E_{1,2} = -4.638191570\,fm^{-1}$, but when $H = 1$, $E_{1,-1} = -4.754198227\,fm^{-1}$ with $\kappa < 0$ and $E_{1,2} = -4.436666340\,fm^{-1}$ with $\kappa > 0$. In summary, the term $2\kappa H$ is responsible for this pseudospin doublet splitting.

To find corresponding wave functions, referring to table 1 and Eqs. (7), we find the functions

$$\rho(s) = s^{-2\sqrt{\Lambda_\kappa(\Lambda_\kappa - 1)C_0 + \frac{\tilde{\gamma}V_3}{4\alpha^2} + \frac{\tilde{\beta}^2}{4\alpha^2}}} (1-s)^{\sqrt{(2\Lambda_\kappa - 1)^2 + \frac{\tilde{\gamma}}{\alpha^2}(V_1 + V_2 + V_3)}} \qquad (22)$$

$$\phi(s) = s^{-\sqrt{\Lambda_\kappa(\Lambda_\kappa - 1)C_0 + \frac{\tilde{\gamma}V_3}{4\alpha^2} + \frac{\tilde{\beta}^2}{4\alpha^2}}} (1-s)^{\frac{1}{2}\left(1 + \sqrt{(2\Lambda_\kappa - 1)^2 + \frac{\tilde{\gamma}}{\alpha^2}(V_1 + V_2 + V_3)}\right)} \qquad (23)$$

$$y_n(s) = P_n^{\left(-2\sqrt{\Lambda_\kappa(\Lambda_\kappa - 1)C_0 + \frac{\tilde{\gamma}V_3}{4\alpha^2} + \frac{\tilde{\beta}^2}{4\alpha^2}},\ \sqrt{(2\Lambda_\kappa - 1)^2 + \frac{\tilde{\gamma}}{\alpha^2}(V_1 + V_2 + V_3)}\right)}(1 - 2s) \qquad (24)$$

By using $G_{n\kappa}(s) = \phi(s)y_n(s)$, we get the lower spinor radial wave functions from Eq. (7) as

$$G_{nk}(s) = B_n s^{-\sqrt{\Lambda_\kappa(\Lambda_\kappa-1)C_0+\frac{\tilde{\gamma}V_3}{4\alpha^2}+\frac{\tilde{\beta}^2}{4\alpha^2}}} (1-s)^{\frac{1}{2}\left(1+\sqrt{(2\Lambda_\kappa-1)^2+\frac{\tilde{\gamma}}{\alpha^2}(V_1+V_2+V_3)}\right)}$$

$$\times P_n^{\left(-2\sqrt{\Lambda_\kappa(\Lambda_\kappa-1)C_0+\frac{\tilde{\gamma}V_3}{4\alpha^2}+\frac{\tilde{\beta}^2}{4\alpha^2}},\sqrt{(2\Lambda_\kappa-1)^2+\frac{\tilde{\gamma}}{\alpha^2}(V_1+V_2+V_3)}\right)}(1-2s) \quad (26)$$

where $B_n$ is the normalized constant. The lower component of Dirac spinor can be calculated by applying Eq. (13b) as

$$F_{n\kappa}(r) = \frac{1}{\left(M - E_{n\kappa} + C_{ps}\right)}\left(\frac{d}{dr} - \frac{\kappa + H}{r}\right)G_{n\kappa}(r) \quad (27)$$

where $E \neq M + C_{ps}$ and only negative energy solutions are valid. The finiteness of our solution requires that the two-components of the wave function be defined over the whole range, $r \in (0,\infty)$. However, in the pseudo-spin limit, if the valence energy is chosen, the upper-spinor component of the wave function will be no longer finite as obviously seen in Eq. (27). Further, introducing the Coulomb-like tensor does not affect the negativity of the energy spectrum in the pseudo-spin limit, but the main contribution is to removing the degeneracy of the spectrum. Of course, the energy eigenvalue equation (21) admits two (negative and positive) solutions (hole and valence states), however, we choose the negative energy solution to make the wave function normalizable in the given range. Aydoğdu and Sever showed that the tensor interaction does not change the radial node structure of the upper and lower components of the Dirac spinor and it effects on the shape of the radial wave functions [29].

On the other hand, to avoid repetition in our solution, the positive valence energy states of Eq. (1), in the spin symmetric case: $\Delta(r) = C_s$, can be easily obtained directly via the pseudospin symmetric solution through the parametric mappings

$$G_{n\kappa}(r) \leftrightarrow F_{n\kappa}(r),\ \kappa \to \kappa+1,\ V(r) \to -V(r),\ E_{n\kappa} \to -E_{n\kappa},\ C_{ps} \to -C_s. \quad (28)$$

In following the previous procedure, one can easily obtain the valence energy eigenvalue equation for the nuclei in the field of the attractive radial potential (1) under the exact spin symmetry limit as

$$\left(2n+1+\sqrt{(2\eta_\kappa-1)^2+\frac{\gamma}{\alpha^2}(V_1+V_2+V_3)}-\sqrt{4\eta_\kappa(\eta_\kappa-1)C_0+\frac{1}{\alpha^2}(\gamma V_3+\beta^2)}\right)^2$$

$$= \frac{\gamma}{\alpha^2}(V_1+V_2+2V_3)+\frac{\beta^2}{\alpha^2}+4\eta_\kappa(\eta_\kappa-1)C_0. \quad (29)$$

where $\gamma = M + E_{n\kappa} - C_s$, $\beta^2 = (M - E_{n\kappa})(M + E_{n\kappa} - C_s)$ and $\eta_\kappa = \kappa + H + 1$. Some numerical results are given in table 3 with parameters set as: $M = 5 fm^{-1}$, $C_s = 0$, $H = 1$, $A = 5$ and $\alpha = 0.6$. Also, we consider the same set of spin doublets as: $(np_{1/2}, np_{3/2})$, $(nd_{3/2}, nd_{5/2})$, $(nf_{5/2}, nf_{7/2})$, $(ng_{7/2}, ng_{9/2})$, …. we see that the tensor interaction removes the degeneracy between two stats in spin doublets. The corresponding wave functions in spin symmetry limit are obtained as

$$F_{n\kappa}(s) = N_{n\kappa} s^{-\frac{1}{2}\sqrt{4\eta_\kappa(\eta_\kappa-1)C_0 + \frac{\gamma V_3}{\alpha^2} + \frac{\beta^2}{\alpha^2}}} (1-s)^{\frac{1}{2}\left(1+\sqrt{(2\eta_\kappa-1)^2 + \frac{\gamma}{\alpha^2}(V_1+V_2+V_3)}\right)}$$
$$\times P_n^{\left(-\sqrt{4\eta_\kappa(\eta_\kappa-1)C_0 + \frac{\gamma V_3}{\alpha^2} + \frac{\beta^2}{\alpha^2}}, \sqrt{(2\eta_\kappa-1)^2 + \frac{\gamma}{\alpha^2}(V_1+V_2+V_3)}\right)}(1-2s), \qquad (30)$$

where $N_{n\kappa}$ is the normalized constant. The lower component of Dirac spinor can be calculated by applying Eq. (13a) as

$$G_{n\kappa}(r) = \frac{1}{M + E_{n\kappa} - C_s}\left(\frac{d}{dr} + \frac{\kappa}{r} - U(r)\right) F_{n\kappa}(r) \qquad (31)$$

where $E \neq -M + C_s$.

### 4. Conclusion

In this work, using the parametric generalization of the NU method, we have obtained approximately the hole energy eigenvalues and the corresponding wave functions of the Dirac equation for the attractive radial potential coupled including a Coulomb-like tensor under the condition of the pseudospin (spin) symmetry limit. Some numerical results are given in table 2 (table 3) and it is shown that the Coulomb-like tensor does not affect the negativity of the energy spectrum in the pseudo-spin (spin) limit, but the main contribution is to removing the degeneracy of the spectrum. In addition, the valence energy states can be produced from our solution for the hole states by taking appropriate transformation of parameters as in (28).

**Table 1.** The specific values for the parametric constants used in calculating the energy eigenvalues and wave functions.

| Constant | Analytical value |
|---|---|
| $c_4$ | $0$ |
| $c_5$ | $-\dfrac{1}{2}$ |
| $c_6$ | $\dfrac{1}{4}\left(1+4\Lambda_\kappa(\Lambda_\kappa-1)C_0+\dfrac{\tilde{\gamma}V_1}{\alpha^2}+\dfrac{\tilde{\beta}^2}{\alpha^2}\right)$ |
| $c_7$ | $\dfrac{1}{4}\left(4\Lambda_\kappa(\Lambda_\kappa-1)(1-2C_0)+\dfrac{\tilde{\gamma}V_2}{\alpha^2}-\dfrac{2\tilde{\beta}^2}{\alpha^2}\right)$ |
| $c_8$ | $\dfrac{1}{4}\left(4\Lambda_\kappa(\Lambda_\kappa-1)C_0+\dfrac{\tilde{\gamma}V_3}{\alpha^2}+\dfrac{\tilde{\beta}^2}{\alpha^2}\right)$ |
| $c_9$ | $\dfrac{1}{4}\left((2\Lambda_\kappa-1)^2+\dfrac{\tilde{\gamma}}{\alpha^2}(V_1+V_2+V_3)\right)$ |
| $c_{10}$ | $1-\sqrt{4\Lambda_\kappa(\Lambda_\kappa-1)C_0+\dfrac{\tilde{\gamma}V_3}{\alpha^2}+\dfrac{\tilde{\beta}^2}{\alpha^2}}$ |
| $c_{11}$ | $2+\left(\sqrt{(2\Lambda_\kappa-1)^2+\dfrac{\tilde{\gamma}}{\alpha^2}(V_1+V_2+V_3)}-\sqrt{4\Lambda_\kappa(\Lambda_\kappa-1)C_0+\dfrac{\tilde{\gamma}V_3}{\alpha^2}+\dfrac{\tilde{\beta}^2}{\alpha^2}}\right)$ |
| $c_{12}$ | $-\dfrac{1}{4}\sqrt{4\Lambda_\kappa(\Lambda_\kappa-1)C_0+\dfrac{\tilde{\gamma}V_3}{\alpha^2}+\dfrac{\tilde{\beta}^2}{\alpha^2}}$ |
| $c_{13}$ | $-\dfrac{1}{2}\left(1+\sqrt{(2\Lambda_\kappa-1)^2+\dfrac{\tilde{\gamma}}{\alpha^2}(V_1+V_2+V_3)}-\sqrt{4\Lambda_\kappa(\Lambda_\kappa-1)C_0+\dfrac{\tilde{\gamma}V_3}{\alpha^2}+\dfrac{\tilde{\beta}^2}{\alpha^2}}\right)$ |

**Table 2.** The bound state energy eigenvalues (in unit of $fm^{-1}$) of the attractive radial potential under pseudospin symmetry limit for several values of $n$ and $\kappa$.

| $\tilde{l}$ | $n, \kappa < 0$ | $(l, j)$ | $E_{n,\kappa<0}$ $H = 0$ | $E_{n,\kappa<0}$ $H = 1$ | $n-1, \kappa > 0$ | $(l+2, j+1)$ | $E_{n,\kappa>0}$ $H = 0$ | $E_{n,\kappa>0}$ $H = 1$ |
|---|---|---|---|---|---|---|---|---|
| 1 | 1, -1 | $1s_{1/2}$ | -4.556531257, 4.685901491 | -4.672750523, 4.849764678 | 0, 2 | $0d_{3/2}$ | -4.556531257, 4.685901491 | -4.352818702, 4.461142207 |
| 2 | 1, -2 | $1p_{3/2}$ | -4.352818702, 4.461142207 | -4.556531257, 4.685901491 | 0, 3 | $0f_{5/2}$ | -4.352818702, 4.461142207 | -4.069044873, 4.166478877 |
| 3 | 1, -3 | $1d_{5/2}$ | -4.069044873, 4.166478877 | -4.352818702, 4.461142207 | 0, 4 | $0g_{7/2}$ | -4.069044873, 4.166478877 | -3.694608131, 3.785571196 |
| 4 | 1, -4 | $1f_{7/2}$ | -3.694608131, 3.785571196 | -4.069044873, 4.166478877 | 0, 5 | $0h_{9/2}$ | -3.694608131, 3.785571196 | -3.201540061, 3.288266427 |
| 1 | 2, -1 | $2s_{1/2}$ | -4.235693135, 4.401472563 | -4.408843996, 4.650285745 | 1, 2 | $1d_{3/2}$ | -4.235693135, 4.401472563 | -3.941172039, 4.073719448 |
| 2 | 2, -2 | $2p_{3/2}$ | -3.941172039, 4.073719448 | -4.235693135, 4.401472563 | 1, 3 | $1f_{5/2}$ | -3.941172039, 4.073719448 | -3.534748813, 3.650186114 |
| 3 | 2, -3 | $2d_{5/2}$ | -3.534748813, 3.650186114 | -3.941172039, 4.073719448 | 1, 4 | $1g_{7/2}$ | -3.534748813, 3.650186114 | -2.986596151, 3.091851290 |
| 4 | 2, -4 | $2f_{7/2}$ | -2.986596151, 3.091851290 | -3.534748813, 3.650186114 | 1, 5 | $1h_{9/2}$ | -2.986596151, 3.091851290 | -2.202461798, 2.301025928 |

**Table 3.** The bound state energy eigenvalues in unit of $fm^{-1}$ of the spin symmetry limit for several values of $n$ and $\kappa$.

| $l$ | $n, \kappa < 0$ | $(l, j = l + 1/2)$ | $E_{n,\kappa<0}$ $H = 1$ | $E_{n,\kappa<0}$ $H = 0$ | $n, \kappa > 0$ | $(l, j = l - 1/2)$ | $E_{n,\kappa>0}$ $H = 1$ | $E_{n,\kappa>0}$ $H = 0$ |
|---|---|---|---|---|---|---|---|---|
| 1 | 0, -2 | $0p_{3/2}$ | -4.964565157 | -4.880113623 | 0, 1 | $0p_{1/2}$ | -4.744442703 | -4.880113623 |
| 2 | 0, -3 | $0d_{5/2}$ | -4.880113623 | -4.744442703, 4.843292115 | 0, 2 | $0d_{3/2}$ | -4.552969050, 4.655853730 | -4.744442703, 4.843292115 |
| 3 | 0, -4 | $0f_{7/2}$ | -4.744442703, 4.843292115 | -4.552969050, 4.655853730 | 0, 3 | $0f_{5/2}$ | -4.298321467, 4.401391558 | -4.552969050, 4.655853730 |
| 4 | 0, -5 | $0g_{9/2}$ | -4.552969050, 4.655853730 | -4.298321467, 4.401391558 | 0, 4 | $0g_{7/2}$ | -3.968507170, 4.071212060 | -4.298321467, 4.401391558 |
| 1 | 1, -2 | $1p_{3/2}$ | -4.859649118 | -4.696712768 | 1, 1 | $1p_{1/2}$ | -4.476756284 | -4.696712768 |
| 2 | 1, -3 | $1d_{5/2}$ | -4.696712768 --- | -4.476756284, 4.630310610 | 1, 2 | $1d_{3/2}$ | -4.189555795, 4.325470989 | -4.476756284, 4.630310610 |
| 3 | 1, -4 | $1f_{7/2}$ | -4.476756284, 4.630310610 | -4.189555795, 4.325470989 | 1, 3 | $1f_{5/2}$ | -3.820028709, 3.947182590 | -4.189555795, 4.325470989 |
| 4 | 1, -5 | $1g_{9/2}$ | -4.189555795, 4.325470989 | -3.820028709, 3.947182590 | 1, 4 | $1g_{7/2}$ | -3.341360523, 3.463125788 | -3.820028709, 3.947182590 |

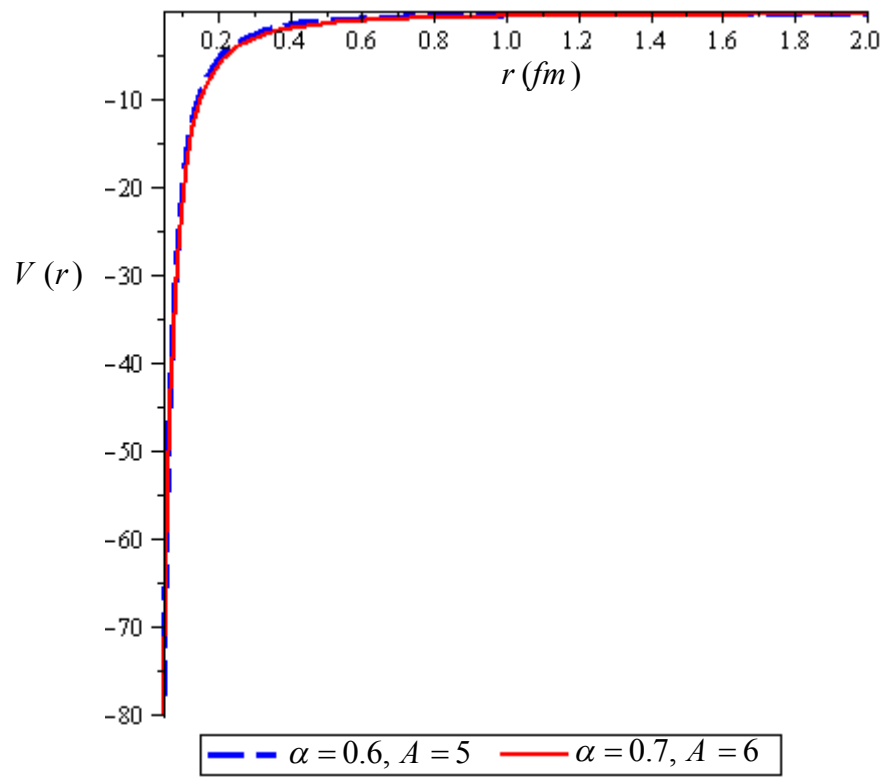

**Figure 1.** Shape of the attractive radial potential.

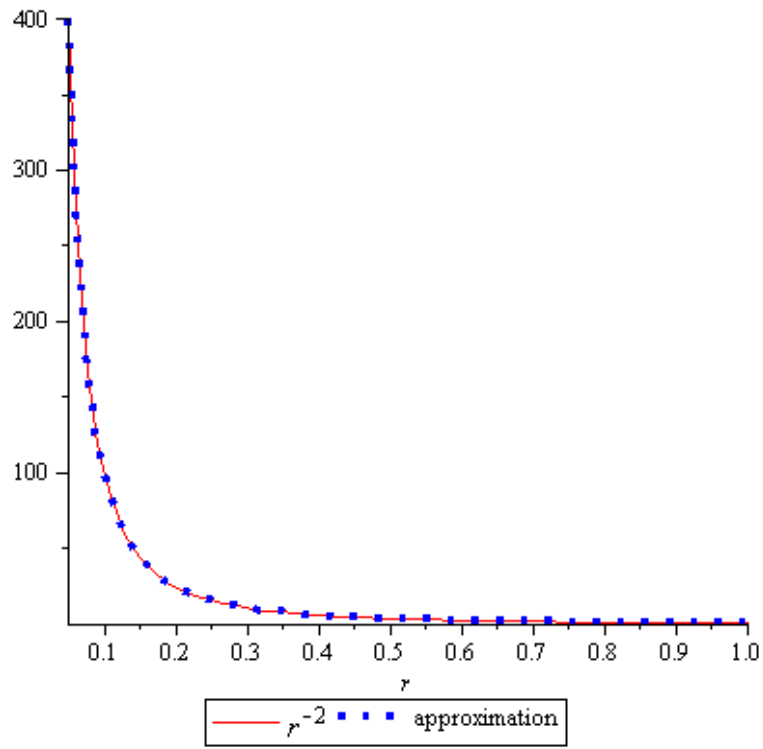

**Figure 2.** Pseudocentrifugal term $\dfrac{1}{r^2}$ (red curve) and its approximation Eq. (18) (blue dot curve).